\documentclass[12pt,preprint]{aastex}

\usepackage{amsmath} 
\usepackage{amsfonts} 
\usepackage{amssymb} 
\usepackage{xspace}
\usepackage{natbib}
\usepackage{wasysym}
\newcommand{\kepler}{$Kepler$\xspace} 	
\newcommand{\teff}{\ensuremath{T_{\mathrm{eff_*}}}\xspace}  
  
\newcommand{\logg}{\ensuremath{\log g}\xspace}                
\newcommand{\z}{\ensuremath{Z}\xspace} 	
\newcommand{\kepmag}{\ensuremath{K_{\mathrm{p}}}\xspace} 	
\newcommand{\msun}{\ensuremath{M_\odot}\xspace}		          
\newcommand{\mstar}{\ensuremath{M_*}\xspace} 		         
\newcommand{\rsun}{\ensuremath{R_\odot}\xspace} 		          
\newcommand{\rstar}{\ensuremath{R_*}\xspace} 		         
\newcommand{\rhostar}{\ensuremath{\bar{\rho_*}}\xspace} 	
\newcommand{\arstar}{\ensuremath{a/R_*}\xspace} 	
\newcommand{\arstarfrac}{\ensuremath{\frac{a}{R_*}}\xspace} 	
\newcommand{\semimaj}{\ensuremath{a}\xspace} 	
\newcommand{\lstar}{\ensuremath{L_*}\xspace} 	
 	
\newcommand{\massp}{\ensuremath{M_{\mathrm{p}}}\xspace} 	
\newcommand{\mjup}{\ensuremath{M_{\mathrm{J}}}\xspace} 	
 	
\newcommand{\rp}{\ensuremath{R_{\mathrm{p}}}\xspace} 	
\newcommand{\rjup}{\ensuremath{R_{\mathrm{J}}}\xspace} 	
\newcommand{\rearth}{\ensuremath{R_{\oplus}}\xspace} 	
\newcommand{\rhop}{\ensuremath{\bar{\rho_{\mathrm{p}}}}\xspace} 	
\newcommand{\teq}{\ensuremath{T_{\mathrm{eq_p}}}\xspace}  
\newcommand{\teffp}{\ensuremath{T_{\mathrm{eff_p}}}\xspace}  
\newcommand{\bimpact}{\ensuremath{b}\xspace} 	
\newcommand{\rprstar}{\ensuremath{R_{\mathrm{p}}/R_*}\xspace} 	
\newcommand{\ted}{\ensuremath{ED}\xspace} 	
\newcommand{\lp}{\ensuremath{L_{\mathrm{p}}}\xspace} 	
\newcommand{\lgen}{\ensuremath{L}\xspace} 	
\newcommand{\rgen}{\ensuremath{R}\xspace} 	
\newcommand{\teffgen}{\ensuremath{T_{\mathrm{eff}}}\xspace} 	
 	
\newcommand{\llambda}{\ensuremath{L_\lambda}\xspace} 	
\newcommand{\blambda}{\ensuremath{B_\lambda}\xspace} 	
\newcommand{\dlambda}{\ensuremath{d\lambda}\xspace}

\newcommand{\gcm}{\ensuremath{\rm{g\,cm^{-3}}}\xspace} 	
 	
\newcommand{\inc}{\ensuremath{i}\xspace} 	
\newcommand{\per}{\ensuremath{P}\xspace} 	
\newcommand{\epoch}{\ensuremath{T_\phi}\xspace} 		         
\newcommand{\chisq}{\ensuremath{\chi^2}\xspace} 		         
\newcommand{\dchisq}{\ensuremath{\Delta \chi^2}\xspace} 		         
\newcommand{\rv}{\ensuremath{K}\xspace} 		         
\newcommand{\dil}{\ensuremath{D}\xspace} 		         
\newcommand{\alb}{\ensuremath{A_{\mathrm{G}}}\xspace} 	
\newcommand{\albb}{\ensuremath{A_{\mathrm{B}}}\xspace} 	
\newcommand{\aell}{\ensuremath{A_{\mathrm{ell}}}\xspace} 
 
\newcommand{\adop}{\ensuremath{A_{\mathrm{dop}}}\xspace}

\newcommand{\sig}{\ensuremath{\sigma}\xspace} 	
\newcommand{\fluxell}{\ensuremath{F_{\mathrm{{ell}}}}\xspace} 		         
\newcommand{\fluxdop}{\ensuremath{F_{\mathrm{{dop}}}}\xspace} 		         
\newcommand{\fluxref}{\ensuremath{F_{\mathrm{{ref}}}}\xspace} 		         
\newcommand{\fluxstar}{\ensuremath{F_*}\xspace} 		         
\newcommand{\fluxtot}{\ensuremath{F_{tot}}\xspace}

\newcommand{\deltaellfrac}{\ensuremath{\frac{\fluxell}{\fluxtot}}\xspace} 	
 	
\newcommand{\deltadopfrac}{\ensuremath{\frac{\fluxdop}{\fluxtot}}\xspace} 	
 	
\newcommand{\deltareffrac}{\ensuremath{\frac{\fluxref}{\fluxtot}}\xspace}

\begin{document} 
\title{Confirmation of Hot Jupiter Kepler-41b via Phase Curve Analysis}

\author{
Elisa~V.~Quintana\altaffilmark{a}, 	
Jason~F.~Rowe\altaffilmark{a},     	
Thomas~Barclay\altaffilmark{b}, 	
Steve~B.~Howell\altaffilmark{c}, 	
David~R.~Ciardi\altaffilmark{d}, 	
Brice-Olivier Demory\altaffilmark{e}, 		
Douglas~A.~Caldwell\altaffilmark{a}, 	
William~J.~Borucki\altaffilmark{c} 		
Jessie~L.~Christiansen\altaffilmark{a}  	
Jon~M.~Jenkins\altaffilmark{a},		
Todd~C.~Klaus\altaffilmark{f} 			
Benjamin~J.~Fulton\altaffilmark{g,h} 	
Robert~L.~Morris\altaffilmark{a}, 		
Dwight~T.~Sanderfer\altaffilmark{c}, 	
Avi~Shporer\altaffilmark{g,i}, 			
Jeffrey~C.~Smith\altaffilmark{a}, 		
Martin~Still\altaffilmark{c}, 			
Susan~E.~Thompson\altaffilmark{a} 	
}
\altaffiltext{a}{SETI Institute, 189 Bernardo Ave, Suite 100, Mountain View, CA 94043, USA}
\altaffiltext{b}{Bay Area Environmental Research Institute/NASA Ames Research Center, Moffett Field, CA 94035, USA}
\altaffiltext{c}{NASA Ames Research Center, Moffett Field, CA 94035}
\altaffiltext{d}{NASA Exoplanet Science Institute/Caltech, 770 South Wilson Ave., MC 100-2, Pasadena, CA 91125, USA} 
\altaffiltext{e}{Department of Earth, Atmospheric and Planetary Sciences, Massachusetts Institute of Technology, 77 Massachusetts Avenue, Cambridge, MA 02139}
\altaffiltext{f}{Orbital Sciences Corporation/NASA Ames Research Center, Moffett Field, CA 94035}
\altaffiltext{g}{Las Cumbres Observatory Global Telescope Network, 6740 Cortona Drive, Suite 102, Santa Barbara, CA 93117, USA}
\altaffiltext{h}{Institute for Astronomy, University of Hawaii, 2680 Woodlawn Dr, Honolulu, HI 96822}
\altaffiltext{i}{Department of Physics, Broida Hall, University of California, Santa Barbara, CA 93106, USA}
\altaffiltext{*}{To whom correspondence should be addressed.  E-mail: elisa.quintana@nasa.gov}

\begin{abstract}
We present high precision photometry of Kepler-41, a giant planet in a 1.86 day orbit around a G6V star that was recently confirmed through radial velocity measurements.  We have developed a new method to confirm giant planets solely from the photometric light curve, and we apply this method herein to Kepler-41 to establish the validity of this technique.  We generate a full phase photometric model by including the primary and secondary transits, ellipsoidal variations, Doppler beaming and reflected/emitted light from the planet.  Third light contamination scenarios that can mimic a planetary transit signal are simulated by injecting a full range of dilution values into the model, and we re-fit each diluted light curve model to the light curve.  The resulting constraints on the maximum occultation depth and stellar density combined with stellar evolution models rules out stellar blends and provides a measurement of the planet's mass, size, and temperature.  We expect about two dozen \kepler giant planets can be confirmed via this method.
\end{abstract}

\clearpage 

\section{Introduction}

NASA's \kepler satellite has been photometrically monitoring more than 150,000 main-sequence stars since its launch in 2009.  The primary goal of the mission is to determine the frequency of Earth-size planets in the habitable zones of Sun-like stars, and in this quest 2326 planetary candidates have been identified with the first 16 months of flight data \citep{2012arXiv1202.5852B}.  Of these, 203 are giant planets with radii (\rp) between 6 -- 15 times the radius of Earth (\rearth).  With \kepler's high photometric precision, both the primary transits and secondary transits (occultations) can be measured for many of these giant planets.  

Occultation measurements allow us to better characterize planets by providing constraints on the size and orbital parameters of a companion that can produce the shape of the light curve.  In addition, depending on the wavelength at which they are gathered, they can provide information to estimate the effective temperature and reflectivity of the planet.  The use of phase curves (the variations in the light curves of a star$+$planet system as the planet orbits the star) as a means to detect exoplanets was presented by \citet{2003ApJ...595..429J}.  Their analysis predicted hundreds of close-in giant planets with periods up to 7 days could be detected by \kepler via their reflected light signatures.  More recent studies have shown that both transiting and non-transiting planets can be detected by measuring the variations in light induced by the companion \citep{2010A&A...521L..59M, 2011MNRAS.415.3921F, 2011AJ....142..195S, 2012MNRAS.422.1512M, 2012A&A...541A..56M}.  Herein, we present a new method to confirm giant planets based solely on the analysis of light curves by modeling both transits and occultations and eliminating other potential non-planetary sources that could produce the shape of the light curve.

Nearby stars that are captured within the target star aperture can dilute the transit signal, resulting in an underestimate of the transit depth and therefore the size of the target star's companion. These third-light or `blend' scenarios can include a background or foreground eclipsing binary star system or a physically bound stellar companion in a hierarchical triple star system \citep{2003ApJ...585.1038S}, each of which has the potential to produce a transit-like signature.  

Confirmation of \kepler candidate planets typically requires additional ground-based follow-up observations due to the prevalence of astrophysical false positives. These techniques, which include spectroscopy, speckle and adaptive optics imaging, precise Doppler measurements and combinations thereof, are used to help eliminate blend scenarios in order to confirm that a \kepler planetary candidate is indeed a planet.  The number of planets that can be confirmed in this manner, however, is limited by the availability of telescope time.  In this article, we present a new method of confirming giant planets without the need for follow-up observations.  We generate a full phase photometric model light curve that includes the primary transits, occultations, ellipsoidal variations, Doppler beaming and reflected/emitted light from the planet. We then inject a full range of dilution values into the model to simulate third-light contamination, then re-fit the diluted light curve models to the photometry.  Comparison of the fitted parameters with stellar evolution models can eliminate systems that are inconsistent with a stellar blend.

To demonstrate this confirmation method, we analyzed the photometry of a star in the \kepler Field of View which shows a signature of a transiting Jupiter-sized planet in a 1.86 day orbit.  Kepler-41b was recently confirmed through radial velocity measurements \citep{2011A&A...536A..70S}, thereby providing a good test case for our method. In Section 2 we present  the photometry of Kepler-41 and discuss our method to correct for systematics and stellar variability.  Our full phase photometric model and best fit parameters are presented in Section 3, and the confirmation technique and results are discussed in Section 4.  Albedo estimates for Kepler-41b and other hot Jupiters are discussed in Section 5, and Section 6 provides a summary.

\section{\kepler Photometry}  
Kepler-41 (RA=19$^{h}$38$^{m}$03$^{s}$.18 and Dec=45$^{\circ}$58'53.9"), also identified by \kepler identification number (KID) 9410930 in the Kepler Input Catalog (KIC) \citep{2011AJ....142..112B} and by \kepler Object of Interest number KOI-196, is a G6V star with an apparent magnitude in the \kepler bandpass of \kepmag = 14.465.  Kepler-41b was identified as a Jupiter-sized candidate companion to this star in \kepler's second data release \citep{2011ApJ...736...19B}.  The SOPHIE spectrograph obtained radial velocity measurements of this object \citep{2011A&A...536A..70S}, which led to an estimated mass  of 0.55 $\pm$ 0.09 \mjup.  

The \kepler observations of Kepler-41 described in this article were acquired between 2009 May 13 and 2011 March 5 and include quarters Q1 through Q8.  Data were sampled nearly continuously during each quarter at 29.42 minute long cadence (LC) intervals (where each LC includes 270 summed 6.5 second exposures).  Although short cadence data (which are sampled more frequently at 58.85 s intervals) are more sensitive to the ingress/egress of a transit and can provide better constraints on the mean stellar density, they are not necessary to measure the phase curve and also are not aways available.  The raw pixels collected for these stars were calibrated \citep{2010SPIE.7740E..64Q} and aperture photometry, background removal and cosmic ray corrections were performed with the Photometric Analysis (PA) software maintained by the \kepler Science Operations Center  to produce the light curves \citep{2010SPIE.7740E..69T}.  Note the light curve data for this target are publicly available at the Mikulski Archive for Space Telescopes (MAST\footnote{http://archive.stsci.edu/kepler}).

The effects of instrumental signals in the flux time series were mitigated by fitting and subtracting cotrending basis vectors\footnote{The cotrending basis vectors are available from \url{http://archive.stsci.edu/kepler/cbv.html}} (their use is documented in \citet{2012MNRAS.422.1219B}) from the light curve using the PyKE software\footnote{PyKE is a suite of Kepler community software available from \url{http://keplergo.arc.nasa.gov/PyKE.shtml}.}. We used the first four cotrending basis vectors and fit them to the data using a linear least-squares approach.  The light curve was stitched together by normalizing the flux by the median value per quarter.  Outliers and additional remaining signatures due to instrumental artifacts (such as those due to `sudden pixel sensitivity dropouts', as described in \cite{2012arXiv1203.1383S}) were then identified and removed.  In total, 215 measurements were discarded yielding a total of 29,680 LC measurements.

We next applied a Fourier decomposition algorithm to separate out star spot-induced variability from the light curve.  Assuming a coherent signal due to the planet and a constant orbital period, sinusoidal components to the light curve were iteratively fit and removed to filter out all frequencies that were not affiliated with the planet orbital period and its associated harmonics.  Specifically, if the amplitude of the peaks in the Fourier Transform were $>$3.6 times the standard deviation (corresponding to 3$\sigma$), we removed those frequencies from the light curve.  The PA-corrected, cotrended, and Fourier-filtered light curves for Kepler-41 are shown in Figure 1. The next section describes our light curve modeling to compute the best-fit planet parameters.

\section{Model Fitting}

Our full phase photometric model uses a circular orbit (eccentricity = 0) and we adopt the formalism of \citet{2002ApJ...580L.171M} to compute the light curve model.  Figure 2 shows the phase-folded and binned photometry for the primary transit (lower curve) centered at orbital phase $\phi$ = 0 with the best-fit model shown in red.  Here, $\phi \equiv 2\pi (t - \epoch)/P$, where  \epoch is the epoch (the time of first mid-transit) and $\it{P}$ is the orbital period of the planet.  The occultation (top curve) near $\phi$ = 0.5 has been magnified (see top and right axes) and the best-fit model is shown in green.  Our model of the transit for Kepler-41 includes nonlinear limb darkening with four coefficients that we compute by performing a trilinear interpolation over \teff, \logg, and \z $\equiv$ [Fe/H] using tables provided by \citet{2011A&A...529A..75C}.  The stellar properties (\teff, \logg, and \z) were adopted from \cite{2011A&A...536A..70S} (shown in Table 1), and the limb darkening coefficients we use are [c1, c2, c3, c4] = [0.6325, -0.3162, 0.8937, -0.4240].  The occultation is modeled in the same manner but we assume the companion is a uniform disk (we neglect limb darkening) due to the relatively short ingress and egress times.  Our model of the phase-dependent light curve takes into account the photometric variability that is induced by the companion, which includes ellipsoidal variations (\fluxell), Doppler beaming (\fluxdop), and light contributed by the planet (\fluxref, which includes both reflected star light and thermal emission).  The relative flux contributions from each of these time-dependent effects are distinct and can be decomposed as 
\begin{eqnarray}	      
F(t) = \frac{\fluxstar + \fluxell(t) + \fluxdop(t) + \fluxref(t)}{\fluxtot}
\end{eqnarray}
where \fluxstar is the illumination measured at phase $\phi$ = 0.5 when the star is blocking the light from the companion, and \fluxtot = \fluxstar + \fluxref($\phi$ = 0).

Ellipsoidal variations in the light curve are caused by changes in the observable surface area of the star due to tidal distortions induced by the companion \citep{2008ApJ...679..783P}.  They have previously been detected in eclipsing binary stars \citep{1976ApJ...203..182W} and more recently in exoplanet systems \citep{2010A&A...521L..59M,2011MNRAS.415.3921F,2011AJ....142..195S,2010ApJ...713L.145W,2012A&A...541A..56M, 2012MNRAS.422.1512M,2012A&A...538A...4M}.  The amplitude of the ellipsoidal variations is roughly equal to the ratio of the tidal acceleration to the stellar surface gravity \citep{2008ApJ...679..783P} assuming tidal equilibrium, and can be approximated by            
   	\begin{eqnarray}	      
	\aell =  \alpha_{ell} \frac{\massp}{\mstar} \left(\frac{\rstar}{a}\right)^3 \rm{sin}^2\it i
	\end{eqnarray}	
where \massp and \mstar are the masses of the planet and star, respectively, \rstar is the stellar radius, \semimaj is the semimajor axis, \inc is the inclination of the orbit relative to the line-of-sight.  The parameter $\alpha_{ell}$ is defined as
	\begin{eqnarray}	      
	\alpha_{ell} = \frac{0.15 (15 + u) (1 + g)}{(3 - u)}
	\end{eqnarray}
	where $u$ and $g$ are the limb darkening and gravity darkening coefficients, respectively \citep{1985ApJ...295..143M}. We compute $u$ = 0.6288 and $g$ = 0.4021 by linearly interpolating over \teff, \logg, and \z using tables provided by \cite{2011A&A...529A..75C}.  For a circular orbit, the contribution of flux from ellipsoidal variations, which oscillate on timescales of half the orbital period (see Figure 3), is
	\begin{eqnarray}	      
	\deltaellfrac = \aell \; \rm{cos}(2\pi\phi \times 2)
	\end{eqnarray}	      
	
Doppler beaming is an apparent increase/decrease in stellar flux due to Doppler shifts in the stellar spectrum that are caused by the reflex star motion around the center of mass due to the companion.  These signals have only recently been measured in transit light curves \citep{2010A&A...521L..59M,2011AJ....142..195S} and oscillate with the orbital period.  Note that we did not detect variations from Doppler beaming in the Kepler-41 light curve, but we include a description here because it may be applicable to other planet candidates.  The amplitude of this signal is   
	\begin{eqnarray}	      	
	\adop = \frac{-\alpha_{dop} 4 K}{c}    
	\end{eqnarray}	      
where $c$ is the speed of light and $\alpha_{dop}$ is a Doppler boosting factor that depends on the wavelength of observation and on the stellar spectrum.  We compute $\alpha_{dop}$ = 1.09 using the methodology as described by \citet{2003ApJ...588L.117L}.  For Keplerian circular orbits, the (non-relativistic) radial velocity semi-amplitude is defined as 
	\begin{eqnarray}	      
	\rv = \left(\frac{2\pi G}{\per}\right)^{1/3} \frac{\massp \rm{sin} \it{\inc}}{\mstar^{2/3}}
	\end{eqnarray}	      
where $\it{G}$ is the gravitational constant.  The contribution of flux from Doppler beaming is
	\begin{eqnarray}	      	
	\deltadopfrac = \adop \; \rm{sin}(2\pi\phi)    
	\end{eqnarray}

The flux variations due to reflected/emitted light from the companion can be approximated by
	\begin{eqnarray}	      
	\deltareffrac = \alb\left({\frac{\rp}{\semimaj}}\right)^2 \Psi(\phi)
	\end{eqnarray}	  
where \alb is the wavelength-dependent geometric albedo, and $\Psi$ is the phase function for a diffusely scattering Lambertian Sphere 
	\begin{eqnarray}	      
	\Psi(\phi) = \frac{\rm{sin}\phi + (\pi - \phi) \rm{cos}\phi}{\pi} 
	\end{eqnarray}

Values of the model parameters are derived using a Levenberg-Marquardt least-squares $\chi^2$ approach \citep{1992nrfa.book.....P}.  We fit for the orbital period (\per), epoch (\epoch), impact parameter (\bimpact), scaled planet radius (\rp/\rstar), geometric albedo (\alb), secondary eclipse depth (\ted), radial velocity semi-amplitude (\rv), ellipsoidal variations (\aell), and mean stellar density (\rhostar).  We then use a Markov Chain Monte Carlo (MCMC) method (eg. Ford 2005) to estimate values and uncertainties using this initial model solution to seed the runs.  Four chains of 10$^6$ samples each are run and the first 25\% are discarded to account for burn in, allowing the Markov chains to stabilize.  The median values of the best-fit parameters for Kepler-41b are given in Table 1, along with the 1\sig (68.3\%) confidence intervals.  We find \massp = 0.598$^{+0.384}_{-0.598}$ \mjup, \rp = 0.996$^{+0.039}_{-0.040}$ \rjup, yielding a mean planet density of \rhop = 0.74$^{+0.48}_{-0.74}$ gm cm$^{-3}$.  The best-fit amplitudes of the occultation depth, ellipsoidal variations, and variations in reflected/emitted light were found to be 60$\pm9$, 4.5$^{+2.8}_{-3.8}$, and 37.4$^{+6.1}_{-6.6}$ ppm, respectively.

\section{Confirmation Method}
Our confirmation method involves two main steps: (1) We first simulate third light contamination scenarios by injecting a wide range of dilution factors into the full phase photometric model and re-fit each diluted light curve model to the photometry.  The results from these model fits set limits on the stellar parameters of possible blends; (2) We next compare these stellar parameters with stellar evolution models to eliminate star/companion systems that are unphysical or inconsistent with a stellar blend.  The goal is to determine the probability that the planet-like signature could be caused by a contaminating star in the aperture.


To model a full range of stellar blends, we iteratively dilute the best-fit model light curve with a dilution factor \dil which ranges from 1\% -- 100\% of the transit depth (using 1\% intervals).  For each value of \dil, we fit the diluted model to the light curve data and recompute \chisq (the goodness-of-fit estimator), \rhostar, \per, \bimpact, \rprstar, \ted, \aell, and \alb.  Figure 4 shows results from these dilution fits for Kepler-41.  The \chisq value and the change in \chisq (\dchisq) as a function of injected dilution are shown in the top two panels.  The lower six panels show the best fit results from six of the above parameters as a function of dilution.  To determine the maximum third light from a potential blend, we solve for the dilution value at which \dchisq increases by 1, 4, or 9, equivalent to 1\sig, 2\sig, or 3\sig (68.3\%, 95.4\%, or  99.7\%) confidence intervals, respectively.  These are shown by red, blue, and green horizontal lines in the top right panel of Figure 4.  This resulted in maximum dilution values of 0.5 (1\sig), 0.6 (2\sig), and 0.67 (3\sig) (shown by the vertical red, blue, and green lines in the lower six panels of Figure 4).  These constraints on the dilution values in turn place limits on the valid parameter values of the companion which we can then compare to stellar evolution models to begin ruling out stellar blends.  

We use Yonsei-Yale (YY) stellar evolution models \citep{2003ApJS..144..259Y,2004ApJS..155..667D} which provide a stellar age, \teff, \rstar, \rhostar, and \logg for a given \mstar and \z.  We start with a grid of stellar masses (\mstar = 0.4 -- 5 \msun with 0.1 \msun increments) and metallicities (\z = 0.00001, 0.0001, 0.0004, 0.001, 0.004, 0.007, 0.01, 0.02, 0.04, 0.06, 0.08) that covers the full range of input values to the YY models.  For each (\mstar, \z) pair, we extract all YY models that have \rhostar values within the constraints set by the dilution fits.  For Kepler-41, the valid values of \rhostar range from 1.17 gm cm$^{-3}$ to a maximum value of 1.7795, 1.7800, or 1.7863 gm cm$^{-3}$ for the 1\sig, 2\sig, and 3\sig constraints, respectively.  Note the confidence intervals that provide constraints on the fit parameters are measured with respect to \dchisq rather than the distribution of each fit parameter.  We require that the age from each model is less than the age of the Universe (taken to be 14 Gyr), and consider only those with $\alpha$-enhanced mixture equal to Solar mixture.  The total number of models extracted for Kepler-41 was 1931, 1947, and 1950 for the 1\sig, 2\sig, and 3\sig cases, respectively.  

For each model, we derive the planet mass \massp, radius \rp, equilibrium temperature \teq, effective temperature \teffp, and mean density \rhop as follows.  We first take the \rhostar value for the given model and interpolate over the valid range of \rhostar (as shown in Figure 4) to find the corresponding dilution factor.  This dilution value is then used to determine the values of the additional fit parameters (\rprstar, \aell, \per, \bimpact, and \ted) for that model.   From these estimates, we can derive the remaining parameters that we use to characterize the star--companion system.

	
The planet radius is found by \rp = (\rp/\rstar)\rstar. The planet mass is roughly proportional to the amplitude of the ellipsoidal variations \citep{2008ApJ...679..783P} and can be estimated using Equation 2.  We solve for \arstar by combining the mean stellar density $\rhostar = \mstar/((4/3)\pi \rstar^3)$ with Kepler's third law (for \massp $\ll$ \mstar)  $\mstar = 4 \pi^2 a^3/(\per^2 G)$, yielding
         	\begin{eqnarray}	      
	\arstarfrac = \left(\frac{\rhostar G \per^2}{3\pi}\right)^\frac{1}{3} 
	\end{eqnarray}
	The inclination is computed from the impact parameter, which (for a circular orbit) is $\bimpact = \arstar\rm{cos}\it{\inc}$.
        The planet mass is then
	\begin{eqnarray}	                      
     	\massp = \frac{(\semimaj/\rstar)^3 \; A_{ell}  \mstar}{\alpha_{ell} \; \rm{sin}^2\it{\inc}}
	\end{eqnarray}
which, along with \rp, provides a measurement of the planet density \rhop.  We note that the amplitude of the Doppler beaming, if detected (which is not the case for Kepler-41), can also be used to estimate the planet's mass \citep{2011AJ....142..195S,2012ApJ...761...53B}.
               
The planet equilibrium temperature can be estimated by
	\begin{eqnarray}	                                       
	\teq = \teff  \sqrt{\frac{\rstar}{2 \semimaj}} \; (f (1- \albb))^{1/4}
 	\end{eqnarray}
where \albb is the wavelength-integrated Bond Albedo (we use \albb = 0.02, which is the approximate value for hot Jupiters in the \kepler bandpass), and $f$ is a circularization factor which equals 1 for isotropic emission \citep{2008ApJ...689.1345R}.  
                
The secondary eclipse depth is approximately equal to the ratio of the planet and star luminosities, $\ted \approx \lp/\lstar$, and can  be used to estimate the planet effective temperature \teffp.  Assuming blackbody radiation, the bolometric luminosity (the total amount of energy emitted across all wavelengths) of an object can be computed from the Stefan-Boltzmann equation $\lgen \propto \rgen^2 \teffgen^4$, where \rgen and \teffgen are the radius and effective temperature of the body, respectively.  The planet effective temperature can be solved from  
	\begin{eqnarray}	                                       
         \ted \approx {\left(\frac{\rp}{\rstar}\right)}^2  {\left(\frac{\teffp}{\teff}\right)}^4  
	\end{eqnarray}
	
Alternatively, we can compute \lp/\lstar by integrating the planet and star Planck functions over the Kepler bandpass ($\lambda$ $\sim$ 400 -- 900 nm).  The wavelength-dependent luminosity can be solved from
	\begin{eqnarray}	                                       
         \llambda \dlambda = 4 \pi^2 \rgen^2 \blambda \dlambda
	\end{eqnarray}
where $\blambda$ is the Planck function: 
	\begin{eqnarray}	                                       
         \blambda = \frac{2hc^2/\lambda^5}{e^{hc/\lambda k \teffgen} - 1}
	\end{eqnarray}
Here, $h$ is Planck's constant, $k$ is Boltzmann's constant and $c$ is the speed of light.  This latter method provides a more accurate estimate of \teffp, and results using both methods to compute \teffp will be discussed in the next section.

\subsection{Results}
Our dilution models at the 1\sig, 2\sig, and 3\sig confidence levels combined with stellar evolution models yield a total of 5828 star/companion model configurations, each providing an estimate of stellar age, \teff, \rstar, \rhostar, \logg, \massp, \rp, \teffp, \teq, and \rhop.  The next step is to examine these models and eliminate those that furnish unphysical stellar properties.  

Figure 5 shows \rhostar as a function of \teff for all available YY evolution models (shown by the red curves in each panel).  Also shown are the values of mean planet density as a function of planet effective temperature (\rhop versus \teffp) for each 3\sig dilution model (shown by the multi-colored tracks in each panel of Figure 5, each color representing a value of metallicity).  In order to comprise a viable stellar blend, a dilution model needs to reside in a region that overlaps with a stellar evolution track.  In the left panel of Figure 5, \teffp was calculated by integrating the Planck function over the Kepler bandpass (as described in the previous section).  None of the resulting dilution models lie in the vicinity of any stellar evolution track, thus eliminating all potential stellar blends in the Kepler-41 photometry.  Using this method to calculate \teffp, we can  conclude that the companion to Kepler-41b is a planet.  

The right panel of Figure 5 shows the dilution models computed using values of \teffp that were estimated from the bolometric luminosities of the star and companion (which are not as precise as those computed by integrating the star and companion Planck functions, but are simpler to derive).  In this case, a subset of the models do overlap with stellar evolution tracks.  Although we have shown that we can use the alternative method of computing \teffp to confirm the planetary nature of Kepler-41b, we show here how we can further examine these overlapping models to exclude them as potential third light contaminants (which may be necessary for confirming other planet candidates).

Figure 6 shows \teq as a function of \teffp for the subset of dilution models that are consistent with stellar evolution tracks (those shown in the right panel of Figure 5).  For the companion to be of planetary nature, we expect a near balance in these temperatures, \teffp $\approx$ \teq, meaning that any incident energy upon the planet is re-radiated.  If \teffp $>>$ \teq, however, the companion must be burning Hydrogen, i.e., of stellar nature (although it is feasible that the object could be a young planet).   For the case of Kepler-41b, \teffp for these remaining dilution models aren't substantially greater than their corresponding values of \teq.  We therefore can't definitively rule out stellar blends using this comparison, but it may be useful for other planet candidate systems.

We next compared the companion mass to the stellar mass (Figure 7) for the same dilution models as shown in Figure 6.  The values of \mstar  are all between 0.8 -- 1.1 \msun whereas the dilution models all have masses below 0.004 \msun, well below the $\sim$0.08 \msun mass limit required for Hydrogen burning \citep{1963ApJ...137.1121K}, i.e., the companion cannot be a star.  With this comparison, we can eliminate the remaining dilution models since we have shown that we cannot produce a proper stellar blend of any kind.  

Our future plans include using this new method of combining phase curve modeling with stellar evolution models to both confirm and characterize additional \kepler planets.  The potential to detect occultations in planet candidate light curves can be determined by combining signal-to-noise measurements with an assumption of an albedo of approximately 30\% (Rowe et al. 2013).  Based on the planet and star characteristics tables from Batalha et al. (2012), we expect about two dozen planet candidates in the \kepler Field of View will have the potential to be confirmed with this method.

\section{The Albedo of Kepler-41b}

\subsection{Hot Jupiters Albedos}
To date, \kepler's unprecedented photometric precision has enabled the detection of planetary occultations in visible wavelengths for eight hot Jupiters: HAT-P-7b \citep{Christiansen:2010a,Welsh:2010}, Kepler-5b \citep{Kipping:2011d,Desert:2011a}, Kepler-6b \citep{Kipping:2011d, Desert:2011a}, Kepler-7b \citep{Kipping:2011d,Demory:2011b}, TrES-2b \citep{Kipping:2011a}, Kepler-12b \citep{Fortney:2011}, Kepler-17b \citep{Desert:2011b,Bonomo:2012} and KOI-196b \citep{Santerne:2011b}.

The {\it Spitzer Space Telescope} gathered thermal planetary emission measurements in infrared wavelengths for several dozens of hot Jupiters. The aforementioned Kepler detections allow us to probe giant planet irradiated atmosphere properties at optical depths that were not explored before, thereby constraining further the energy budget of hot Jupiter planets \citep[e.g.,][]{Madhusudhan:2009c}.

In our solar system, gas giant geometric albedos range from 0.32 for Uranus to 0.50 for Jupiter in a bandpass similar to \textit{Kepler}'s \citep{Karkoschka:1994}. This is mainly due to their low equilibrium temperatures, as compared to hot Jupiters, which allow the formation of cloud decks made of ammonia and water ice in their atmosphere that are highly reflective in visible wavelengths \citep{2011ApJ...735L..12D}.

Hot Jupiters emit very little in visible wavelengths.  The albedos of hot Jupiters were expected to be low due to efficient reprocessing of stellar incident radiation into thermal emission \citep{Marley:1999,Seager:2000a,Sudarsky:2003}. In addition, the presence of alkali metals in hot Jupiter atmospheres (Na and K) as well as TiO and VO (at the hotter range) is expected to cause significant absorption at visible wavelengths, rendering most hot Jupiters dark.

The first constraint on a hot Jupiter visible flux was obtained with the MOST satellite \citep{Walker:2003} observing HD209458b \citep{Rowe:2008}. The corresponding geometric albedo 3$\sigma$ upper-limit of $A_g < 0.08$ confirmed these earlier theoretical predictions. The majority of hot Jupiter occultations measured by {\it Kepler} photometry corroborate today the hypothesis  that hot Jupiters emit very little in visible wavelengths, their measured geometric albedo being attributed to thermal emission leaking into shorter wavelengths rather than contribution from Rayleigh scattering, clouds or hazes. 

Remarkably, a few irradiated giant planets exhibit visible flux in the  {\it Kepler} bandpass that exceeds the expected contribution from thermal emission alone. A recent detailed analysis of Kepler-7b occultation measurements showed a significant departure of the measured brightness temperature as compared to the equilibrium temperature, suggesting that the planetary flux is dominated by Rayleigh scattering and/or hazes \citep{Demory:2011b}. In addition, combining visible and {\it Spitzer} infrared occultation measurements showed that Kepler-12b also exhibits an excess of flux in the visible, possibly indicating a reflective component in this low-density hot Jupiter atmosphere \citep{Fortney:2011}.  Ideas that have been invoked to explain the wide variation in observed hot Jupiter albedos include variations in planetary densities \citep{Sudarsky:2003} and condensates phase transitions at narrow temperature ranges \citep{Demory:2011b,Kane:2012a}.

\subsection{Kepler-41b as another outlier?}
Our global analysis yields an occultation depth of $60\pm9$ ppm, which translates to a geometric albedo of $A_g = 0.23\pm0.05$. Using a blackbody  spectrum for the host star, the corresponding brightness temperature is 2420 K, which is $\sim$400 K larger than the maximum planetary equilibrium temperature, assuming zero-Bond albedo and no stellar incident energy recirculation from the day hemisphere to the night hemisphere.  Kepler-41b shows similar brightness temperature excess as Kepler-7b, possibly suggesting contribution from Rayleigh scattering and/or hazes.

The planetary phase modulation, caused by the combination between reflected light and thermal emission, has an amplitude that is $\sim$1\sig smaller than the occultation depth and slightly offset from the mid-occultation timing. At the high atmospheric pressures probed by \kepler (P$\sim$1 bar), we would expect the even temperature across hemispheres to yield a phase curve exhibiting only nominal modulation. This result suggests that either atmosphere dynamics at depth are significantly more complex than this description or that the phase curve modulation is dominated by reflected light instead of thermal emission.  Detailed modeling and {\it Spitzer} infrared observations would be especially useful toward a precise constraint on the planetary energy budget and could unambiguously disentangle the thermal emission and reflected light components.

\section{Summary} 
We have presented a new method to confirm giant planets purely by analysis of the photometric light curve combined with stellar evolution models.  We have developed a full phase photometric model that includes both primary and secondary transits along with flux contributions from ellipsoidal variations, Doppler beaming, and reflected/emitted light. 
We inject a full range of dilution values into the model light curve to simulate third light contamination from stellar blends, and iteratively fit each diluted model light curve to the photometry.  We then compare these fit results to stellar evolution models to determine if any set of diluted model parameters are valid (meaning the star and companion have  masses, sizes and orbits that are consistent with a stellar evolution model) and match the shape of the photometric light curve.

We applied this method to Kepler-41, a G6V star with a recently confirmed giant planet \citep{2011A&A...536A..70S}, using \kepler photometry taken during quarters Q1 -- Q8.  The phased light curve shows a clear secondary occultation with a depth of 60$\pm9$ ppm.  The phase of this occultation is near $\phi$ = 0.5, indicating the orbit of Kepler-41b is likely nearly circular.   We detected flux variations due to reflected/emitted light from the planet (with an amplitude of 37.4$^{+6.1}_{-6.6}$ ppm) and ellipsoidal variations  (4.5$^{+2.8}_{-3.8}$ ppm), the latter of which enables us to estimate the mass and density of the planet.  We did not detect variations due to Doppler beaming, but these measurements -- if detected in the light curves of other planetary candidates -- can also be used to measure the planet mass. 

To determine whether any dilution models have properties consistent with a star, we first compared the denities and effective temperatures derived from the diluted models (\rhop versus \teffp) to all Yonsei-Yale evolution tracks (which provide \rhostar as a function of \teff for all valid stellar evolution models).  To estimate \teffp for the dilution models, we computed the ratio of the planet luminosity to that of the star (\lp/\lstar), which is approximately equal to the measured secondary eclipse depth, and solved for \teffp.  We first computed \lp/\lstar using the bolometric luminosities (the total amount of energy emitted across all wavelengths) for the planet and star.  For comparison, we also computed these luminosities by integrating the Planck functions over the Kepler bandpass.  This latter method to compute \teffp resulted in unphysical star/companion parameters for all dilution models, thereby eliminating the possibility that the companion to Kepler-41b could be a stellar blend.  Using values of \teffp that were computed from bolometric luminosities yielded a small subset of dilution models that were consistent with stellar evolution tracks.  For these models, we further examined the temperatures and masses of each system to filter out additional inconsistencies.  We found that all companion masses from these remaining diluted models were well below the $\sim$0.08 \msun limit for Hydrogen-burning, indicating that the companion cannot be a star.  Although both methods to compute \teffp provided enough information to rule out stellar blends in the Kepler-41 photometry, we recommend computing \teffp with the more accurate method  of integrating the star and planet luminosities over the Kepler bandpass.

Our best-fit model of Kepler-41b yields \massp = 0.598$^{+0.384}_{-0.598}$ \mjup and \rp = 0.996$^{+0.039}_{-0.040}$ \rjup.  From our analysis of the phase curve combined with stellar evolution models we can therefore independently confirm that Kepler-41b is indeed a planet.  This confirmation method can be applied to additional \kepler planet candidates that show a clear occultation in their light curve. 

%

\acknowledgements
This paper includes data collected by the Kepler mission. Funding for the Kepler mission is provided by the NASA Science Mission directorate. Some/all of the data presented in this paper were obtained from the Mikulski Archive for Space Telescopes (MAST). STScI is operated by the Association of Universities for Research in Astronomy, Inc., under NASA contract NAS5-26555. Support for MAST for non-HST data is provided by the NASA Office of Space Science via grant NNX09AF08G and by other grants and contracts.


\clearpage
\begin{deluxetable}{lllll}
\tablecolumns{5} 
\tabletypesize{\scriptsize} 
\tablecaption{Star and Planet Parameters for Kepler-41} 
\label{tbl-2}
\tablewidth{0pt} 
\tablehead{
\colhead{$\bf Parameters$} &
\colhead{$\bf Med$} &
\colhead{$\bf Std$} &
\colhead{$\bf +1\sigma$}   &
\colhead{$\bf -1\sigma$}}
\startdata 


\mstar (\msun)  & 1.044    &      0.062    &      0.059    &     -0.068  \\ 
\rstar (\rsun)  &  1.065    &      0.040    &      0.042     &    -0.042  \\ 
\z                &  0.036     &     0.011     &     0.010     &    -0.012      \\ 
\teff (K)     &  5617    &       143      &     126     &     -177   \\ 
log \lstar ($L_\odot$)     & 0.007      &    0.064    &      0.062    &     -0.069   \\ 
\logg (cgs)&  4.398    &      0.028     &     0.027     &    -0.032  \\ 
\rhostar (\gcm) & 1.213    &      0.099      &    0.081   &      -0.113 \\

 & & & & \\

Period (days)    & 1.8555577   &   0.0000003   &   0.0000003   &  -0.0000004   \\ 
T$_0$ (BJD-2454900) & 70.180267   &    0.000067    &   0.000065  &    -0.000070  \\ 
b                   &  0.475   &       0.048    &      0.045     &    -0.044  \\ 
\rp/\rstar        & 0.09601   &     0.00070   &     0.00070    &   -0.00070 \\ 
a/\rstar             & 6.04    &       0.16     &      0.14    &      -0.18     \\ 
\rp (\rjup)  &  0.9956    &       0.0375  &       0.0393       &   -0.0401  \\ 
a (AU)              &   0.02999   &     0.00060    &    0.00052   &    -0.00071  \\ 
\massp (\mjup)  & 0.5978    &      0.7803    &      0.3839   &      -0.5978 \\ 

\rhop (g/cm$^3$) &0.74      &     0.98     &      0.48     &     -0.74    \\ 
\inc (deg)        & 85.50     &      0.47      &     0.40     &     -0.50      \\ 
Duration (h)         &  2.335     &     0.015     &     0.015    &     -0.016  \\   
Depth (ppm)     &   10653       &     13      &      13      &     -13    \\ 

 & & & & \\

Occultation depth (ppm)&  60      &      11    &6       9      &    -9    \\ 
Ellipsodial (ppm)&   4.5      &      3.3     &       2.8    &       -3.8   \\  
Reflected/Emitted (ppm)     & 37.4    &        6.3      &      6.1     &      -6.6    \\ 
K (m/s)         &   94     &      123       &     61       &    -94  \\ 

 & & & & \\
\enddata 
\end{deluxetable}

\clearpage

\begin{figure}
\plotone{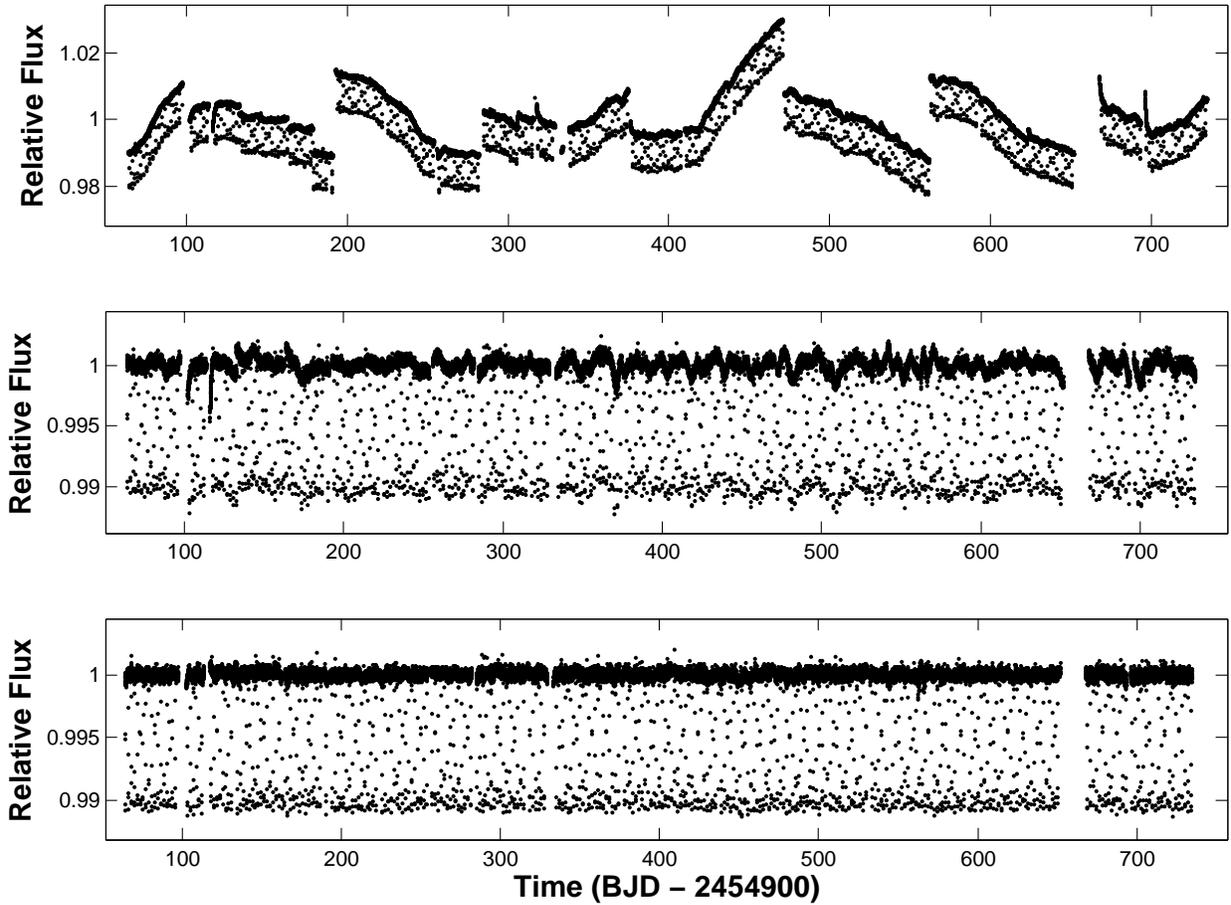}
\figcaption{\kepler Q1 -- Q8 PA photometry for Kepler-41 is shown in the top panel.  The cotrended light curve (middle panel) shows mostly spot activity.  The light curve was Fourier-filtered to remove stellar variability (lower panel).}
\end{figure}

\begin{figure}
\plotone{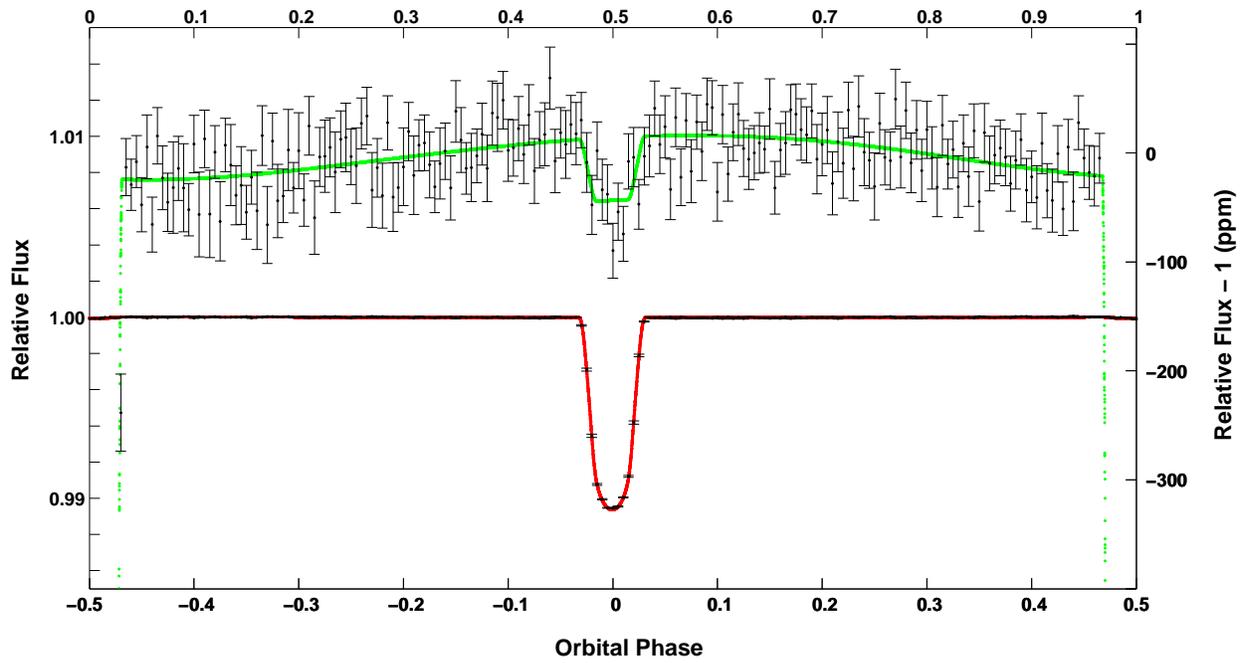}
\figcaption{The detrended data for Kepler-41 phased to the orbital period and binned to 0.005 in phase. The red-lined data are centered on the transit and show the full phase and the green-lined data are centered on the occultation and magnified (see top and right axes).  The green curve fits for ellipsoidal variations, Doppler boosting, and reflected/emitted light.}
\end{figure}

\begin{figure}
\plotone{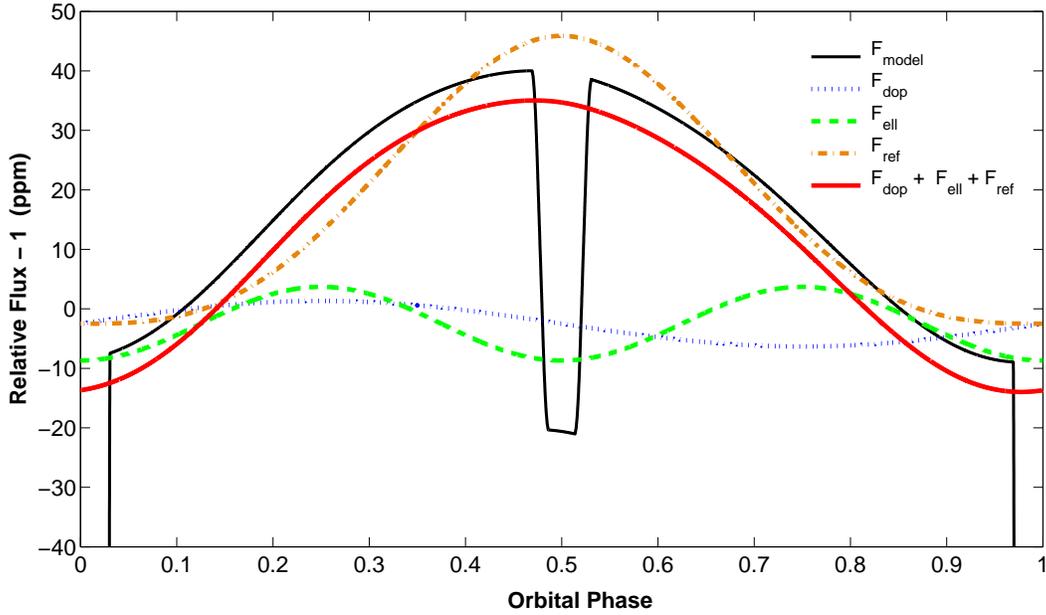}
\figcaption{The best-fit model for Kepler-41b phased to the orbital period and magnified to show the occultation.  Our full phase photometric model includes flux variations induced by the companion that can be decomposed.  These include Doppler beaming (blue dotted curve), ellipsoidal variations (green dashed curve) and reflected/emitted light (orange dot-dashed curve).  The sum of these three effects is shown in red.  Note we did not detect Doppler beaming in the light curve of Kepler-41, but we include a description of this effect in this article because it may be applicable to other planet candidates.}
\end{figure}

\begin{figure}
\epsscale{0.70} 
\plotone{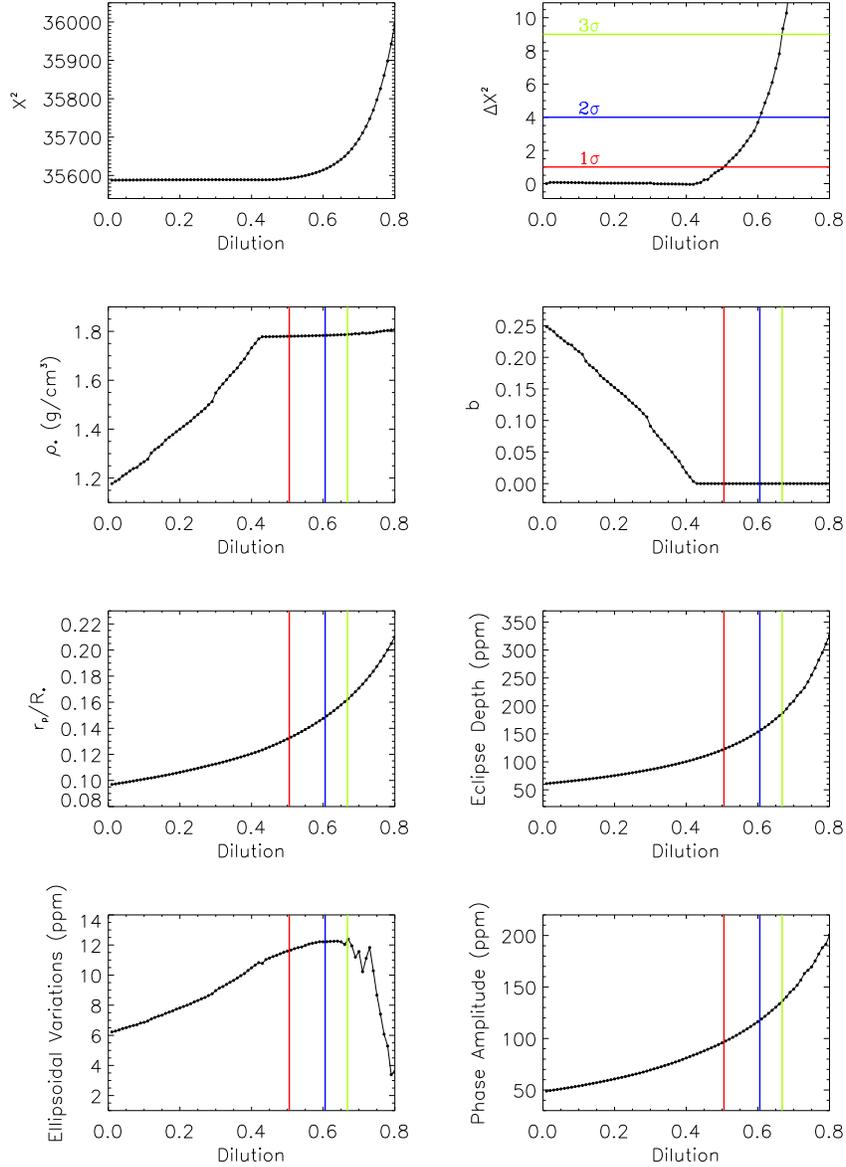}
\figcaption{Results from our dilution model fits. The goodness-of-fit estimator \chisq is shown in the top left panel as a function of dilution values that were injected into the light curve.  We solve for the maximum allowed dilution (i.e., the maximum amount of 3rd light from a potential blend) by measuring where \dchisq changes by 1, 4, or 9 (corresponding to 1\sig, 2\sig, or 3\sig), as shown in the top right panel by the red, blue and green horizontal lines, respectively.  The lower six panels show six of the fit parameters as a function of dilution, and the red, blue and green vertical lines determine their range of valid values as constrained by the dilution fits.  Comparison of each valid dilution model to stellar evolution models rules out massive, stellar objects, confirming the planetary nature of Kepler-41b.}
\end{figure}

\begin{figure}
\epsscale{1.01} 
\plotone{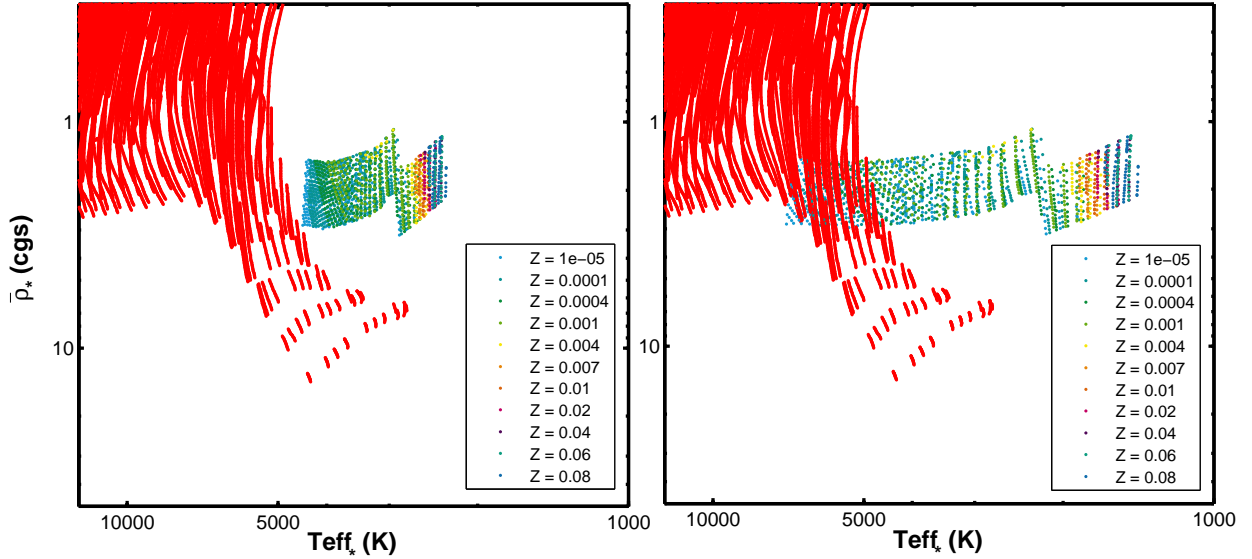}
\figcaption{The mean stellar density \rhostar is shown here as a function of \teff for all available Yonsei-Yale stellar evolution tracks (shown by the red curves in each panel).  The companion \rhop and \teffp from the dilution model fits are overplotted for a range of metallicities \z (colored points in each panel).  The dilution models in the left panel were computed using estimates of \teffp that were computed by integrating the planet and star Planck functions over the Kepler bandpass and comparing the ratio of the resulting luminosities to the secondary eclipse depth \ted.  All dilution models in this case are inconsistent with any stellar blend (there is no overlap with the stellar evolution tracks), and we can conclude that the companion to Kepler-41 is a planet.  In the right panel, the dilution models were computed using \teffp values that were calculated from the ratio of the planet and star bolometric luminosities (over all wavelengths).  This was done to determine if this simpler method (albeit not as precise) to compute \teffp is sufficient to rule out potential blends. In this case, a subset of dilution models overlap with stellar evolution tracks and therefore need to be examined further (see Figures 6 and 7) in order to rule out stellar blends.}
\end{figure}

\begin{figure}
\epsscale{0.9} 
\plotone{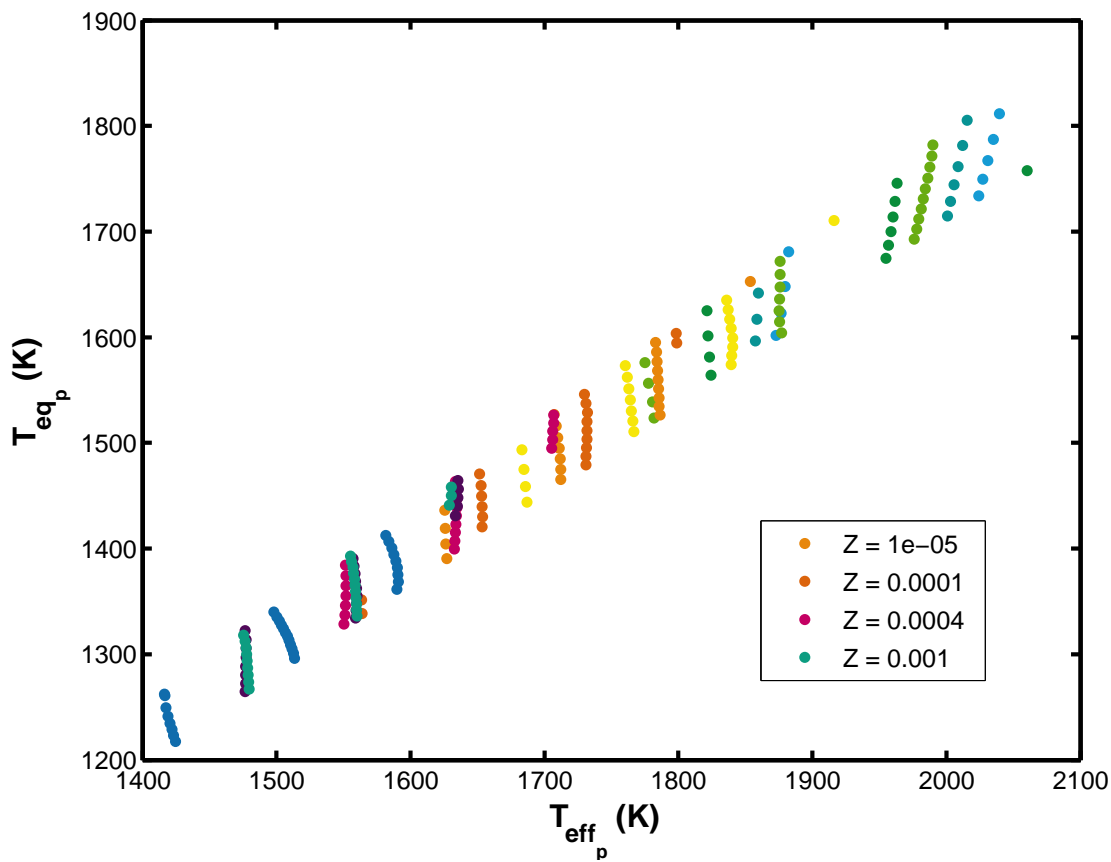}
\figcaption{For the remaining valid dilution models (those that have parameters that overlap with stellar evolution tracks as shown in Figure 5), the equilibrium temperatures can be compared to the effective temperatures.  To be of stellar nature, the values of \teffp for each model would need to be much greater than the corresponding values of \teq (indicating that the companion is burning Hydrogen). In this case, the temperatures are comparable and cannot be used to definitively rule out stellar blends, but this comparison may be useful to confirm other planet candidates.}
\end{figure}

\begin{figure}
\epsscale{0.90} 
\plotone{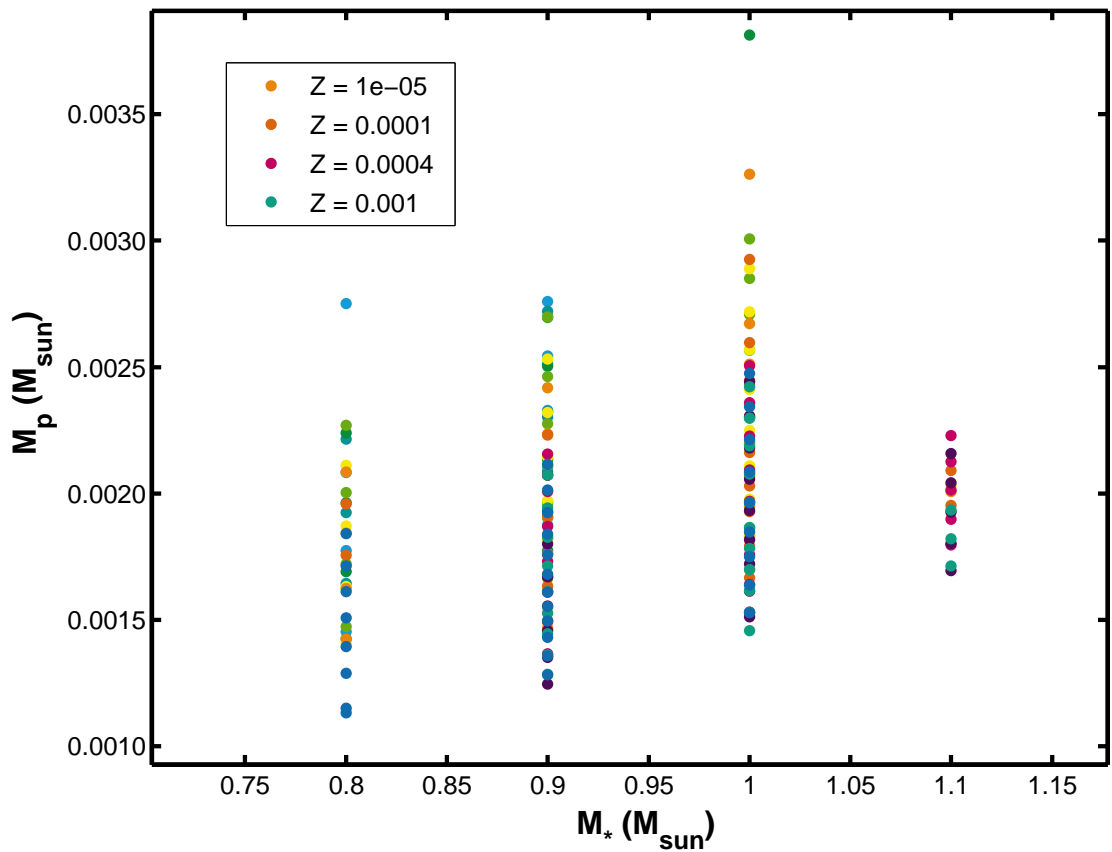}
\figcaption{For the remaining valid dilution models (as shown in Figure 6), the relation between each companion mass \massp and the corresponding stellar mass \mstar is shown here.  All dilution models have a companion mass less than that needed for Hydrogen burning ($\sim$0.08 \msun),  indicating that the companion cannot be a star.  With this comparison, we can eliminate these remaining dilution models and conclude that the companion to Kepler-41b is a planet.}
\end{figure}

\clearpage

\bibliographystyle{apj}
\bibliography{references}

\end{document}